\documentclass[aps,pre,onecolumn,groupedaddress]{revtex4}

\usepackage{graphicx,psfrag}
\usepackage{amsmath}

\def\BibTeX{{\rm B\kern-.05em{\sc i\kern-.025em b}\kern-.08em
    T\kern-.1667em\lower.7ex\hbox{E}\kern-.125emX}}

\begin{document}

\title{Measure for degree heterogeneity in complex networks and its application to recurrence network analysis}

\author{Rinku Jacob}
\email{rinku.jacob.vallanat@gmail.com}
\affiliation{Department of Physics, The Cochin College, Cochin-682 002, India}
\author{K. P. Harikrishnan}
\email{kp_hk2002@yahoo.co.in}
\affiliation{Department of Physics, The Cochin College, Cochin-682 002, India} 
\author{R. Misra}
\email{rmisra@iucaa.in}
\affiliation{Inter University Centre for Astronomy and Astrophysics, Pune-411 007, India} 
\author{G. Ambika}
\email{g.ambika@iiserpune.ac.in}
\affiliation{Indian Institute of Science Education and Research, Pune-411 008, India} 

\begin{abstract}
We propose a novel measure of degree heterogeneity, for unweighted and undirected complex networks, which   
requires only the degree distribution of the network for its computation. We show that 
the proposed measure can be applied to all types of network topology with ease and increases with the diversity 
of node degrees in the network.   
The measure is applied to compute the heterogeneity of synthetic (both random and scale free) 
and real world networks with its value normalized in the interval $[0,1]$.   
To define the measure, we introduce a limiting network whose heterogeneity can be expressed analytically 
with the value tending to 1 as the size of the network $N$ tends to infinity. We numerically study 
the variation of heterogeneity for random graphs (as a function of $p$ and $N$) and for scale free 
networks with $\gamma$ and $N$ as variables.  
Finally, as a specific application, we show that the proposed measure can be  
used to compare the heterogeneity of recurrence networks constructed from the 
time series of several low dimensional chaotic attractors, thereby providing a single index to 
compare the structural complexity of chaotic attractors.    
\end{abstract}

\maketitle


\section{\label{sec:level1}INTRODUCTION}
A network is an abstract entity consisting of a certain number of nodes connected by links or 
edges. The number of nodes that can be reached from a reference node $\imath$ in one step is called its 
degree denoted by $k_i$. If equal number of nodes can be reached in one step from all the nodes, the network 
is said to be regular or homogeneous. A regular lattice where nodes are associated with fixed locations in 
space and each node connected to equal number of  nearest neighbours, is an example of a regular network. 
However, in the general context of complex networks, it is  
defined in an abstract space with a set of nodes $\mathcal N = {1,2,3....N}$ and a set of links denoted by 
$\mathcal K = {k_1,k_2,k_3 ....k_{N-1}}$. As the spectrum of $k$ values of the nodes increases, the network 
becomes more and more irregular and complex. Over the last two decades, the study of such complex networks 
has developed into a major field of inter-disciplinary research spanning across mathematics, physics, biology 
and social sciences \cite {new1,barz,new2}.

Many real world structures \cite {alb1} and interactions \cite {barz,mil} can be modeled using the 
underlying principles of complex networks and analysed using the associated network measures \cite {boc}. 
In such contexts, the corresponding complex 
network can be weighted \cite {yoo} or unweighted and directed \cite {barr} or undirected depending on 
the system or interaction it represents. In this paper, we restrict ourselves to unweighted and undirected 
networks and the possible extensions for weighted and directed networks are discussed in the end.  
The topology or structure of a complex network is determined by the 
manner in which the nodes are connected in the network. For example, in the case of the classical random 
graphs (RG) of Erd\H os and R\' enyi (E-R) \cite {erd},  
two nodes are connected with a constant and random probability $p$. In contrast, many real world networks 
are found to have a tree structure with the network being a combination of small number of \emph {hubs} on to 
which large number of individual nodes are connected \cite {cal}. An important measure that distinguishes 
between different topologies of complex networks is the degree distribution $P(k)$ that determines how many 
nodes in the network have a given degree $k$. For the RGs, $P(k)$ is a Poisson distribution around the 
average degree $<k>$ \cite {boc} while many real world networks follow a fat-tailed power law distribution 
given by $P(k) \propto k^{-\gamma}$, with the value of $\gamma$ typically between 1 and 3 \cite {bar1}. 
Such networks are called \emph {scale free} (SF) \cite {alb2,est1} due to the inherent scale invariance of 
the distribution. 

Though topology is an important aspect of a complex network, that alone is not sufficient to characterize 
and compare the interactions that are so vast and diverse. A number of other statistical measures have been 
developed for this purpose, each of them being useful in different contexts. Two such commonly used quantifiers 
are the clustering coefficient (CC) and the characteristic path length (CPL). There are also characteristic 
properties of local 
structure used to compare the complexity of networks in particular cases, such as, the hierarchy or community 
structure \cite {new2} in social networks and motifs \cite {xu} and super family profiles \cite {mil} in 
genetic and neuronal networks. However, a single index that can quantify the diversity of connections between 
nodes in  networks even with different topologies, is the heterogeneity measure \cite {est2}. It is also 
indicative, in many cases, of how stable and robust \cite {cruc} a network is with respect to perturbations from various 
external parameters. An important example is the technological network of North American power grid \cite {alb1}. 
Recent studies have also revealed the significance of the heterogeneity measure in various other contexts, such as, 
epidemic spreading \cite {past}, traffic dynamics in networks \cite {zhao} and network synchronization \cite {mott}.  

The network heterogeneity has been defined in various ways in the literature which we will discuss in detail in 
the next section where, we will also present the motivations and need for a new measure. While all the 
existing measures are based on the degree correlations $k_i$ and $k_j$ of nodes 
$\imath$ and $\jmath$ in the network, the measure proposed in this paper uses only the degree distribution $P(k)$ to 
compute the heterogeneity of the network. However, we show that this new measure varies directly with the 
$k$ spectrum, or the spectrum of $k$ values in the network, and hence gives a true representation of the 
diversity of node degrees present in the network. In other words, it serves as a single index to quantify the 
node diversity in the network. 

In this work, we also include  a class of networks not considered so far in the context of heterogeneity measure 
in any of the previous works. These are complex networks constructed from the time series of chaotic dynamical systems, 
called recurrence networks \cite {don1}. They have a wide range of practical applications \cite {mar1,avi} and the 
measures from these networks are used to characterize strange attractors in state space, typical of chaotic 
dynamical systems, as discussed in \S V. The diversity of node degrees in the RNs was actually one of the 
motivations for us to search for a heterogeneity measure that could be used to compare the structural 
complexities of different chaotic attractors through the construction of RNs.

Our paper is organized as follows: In the next section, we discuss briefly all the previous measures of 
heterogeneity and give reasons why we have to look for a new measure. The measure that we propose is based on 
the idea of what we consider as a completely heterogeneous network of $N$ nodes, that is illustrated in \S III. 
The proposed measure of heterogeneity is presented in \S IV while \S V and \S VI are devoted to computation of this 
new measure for various synthetic as well as real world networks. Our conclusions are summarised in \S VII.

\section{\label{sec:level1}EXISTING MEASURES OF HETEROGENEITY}
If we carefully analyze the heterogeneity measures proposed in the literature, it becomes clear that 
two different aspects of a complex network can be quantified through a heterogeneity measure. They are the 
diversity in node degrees and the diversity in the structure of the network. For example, the 
initial attempts to measure the heterogeneity try to capture the diversity in the node degrees of the network  
and were mainly motivated by 
the random graph theory. The first person to propose a measure of heterogeneity was Snijders \cite {sni} 
in the context of social networks and it was modified by Bell \cite {bell} as the variance of node degrees:
\begin{equation}
VAR = {{{1} \over {N}} \sum_{i}^N (k_i - <k>)^2}
  \label{eq:1}
\end{equation} 
where $<k>$ represents the average degree in the network. Though this is still one of the popular measures of 
heterogeneity, its applicability is mainly limited to RGs where one can effectively define an average $k$. 
Another measure was proposed by Albertson \cite {bert} as:
\begin{equation}
A = \sum_{i,j} |k_i - k_j|
  \label{eq:2}
\end{equation} 
which is a sum of the local differences in the node degrees in the network.  
This index also is not completely adequate in quantifying correctly the heterogeneity of networks with 
different topologies. Apart from the above two measures defined in the context of social networks, another 
measure \cite {huwa} has recently been proposed to quantify the degree heterogeneity. It uses a measure of 
inequality of a distribution, called the Gini coefficient \cite {cola}, which is widely used in economics 
to describe the inequality of wealth. Here a heterogeneity curve is generated using the ratio of cumulative 
percentage of the total degree of nodes to the cumulative percentage of the number of nodes. The heterogeneity 
index is then measured as the degree inequality in a network. Though the authors compute heterogeneity of 
several standard exponential and power law networks, the measure turns out to be very complicated and works 
mainly for networks of large size with $N \rightarrow \infty$. In short, none of these measures, though useful 
in particular contexts, truly reflects  heterogeneity as represented by the diversity of node degrees in a network.
A comparative study of the above heterogeneity measures has been done by Badham \cite {badham}. 

The second aspect of heterogeneity discussed in the literature 
is the topological or structural heterogeneity possible in a complex network which is 
especially important in real world networks. An example for this is the measure proposed by Estrada \cite {est2} 
recently, given by 

\begin{equation}
\rho = \sum_{i,j} ({{1} \over {\sqrt k_i}} - {{1} \over {\sqrt k_j}})^2
  \label{eq:3}
\end{equation} 
which can also be normalised to get a measure $\rho_n$ within the unit interval $[0,1]$ as:

\begin{equation}
\rho_n = {{\rho} \over {N - 2 \sqrt (N-1)}}
  \label{eq:4}
\end{equation} 

\begin{figure}
%\centering
\includegraphics[width=0.9\columnwidth]{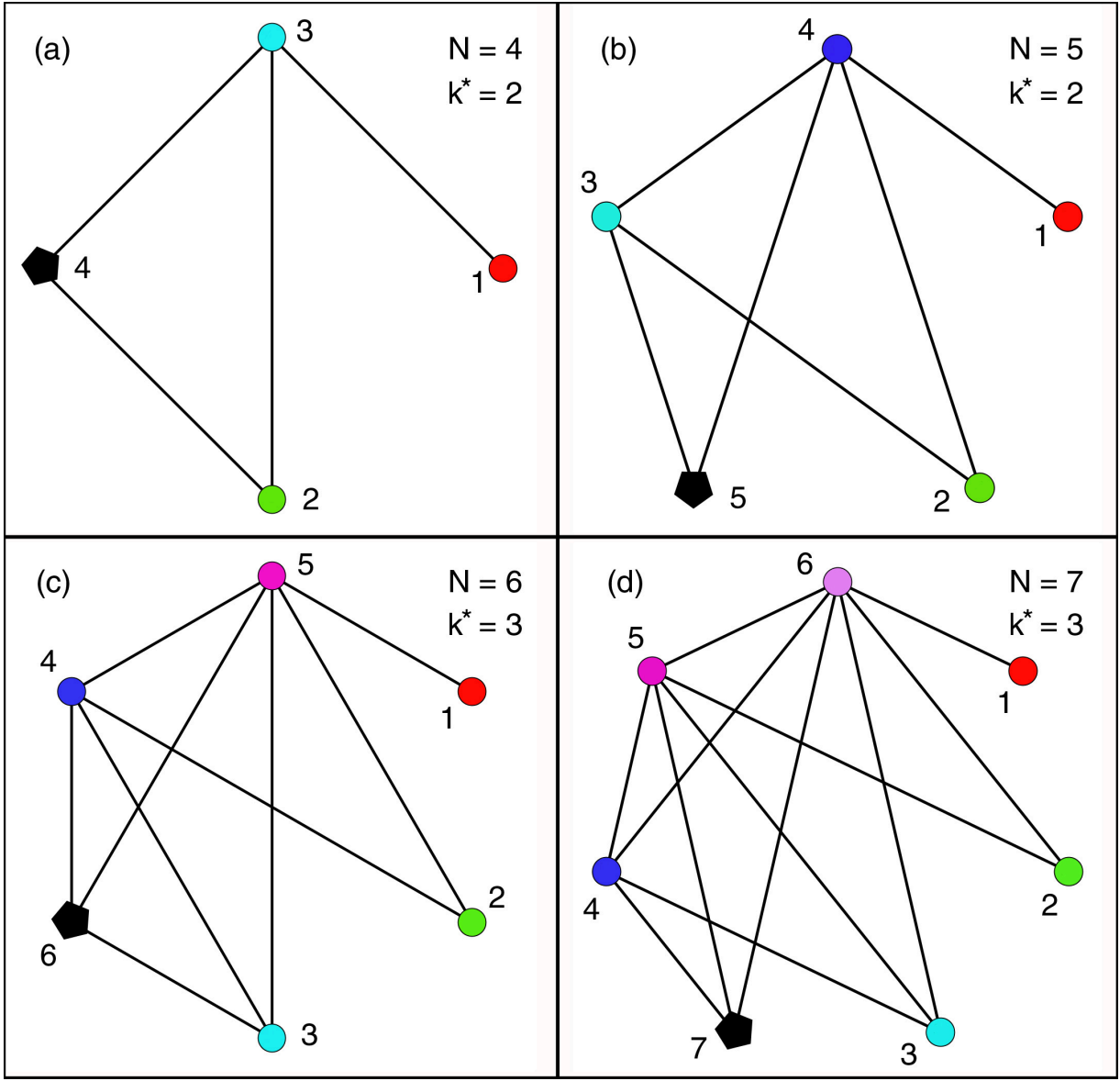}%
\caption{\label{f.1}A comparison of the completely heterogeneous networks (see text) with 
$N = 4$, $5$, $6$ and $7$. In each case, all the possible $k$ - values from $1$ to $(N-1)$ are present 
in the network as shown. One degree (one $k$ value) has to be shared by two nodes since the $N^{th}$ 
node will have the degree of any one of other nodes. It is empirically shown that this degree of 
$N^{th}$ node, denoted by $k^{\ast}$, is automatically fixed (if the network has all possible degrees from $1$ to 
$(N-1)$) and is $N/2$ if $N$ is even and $(N-1)/2$ 
if $N$ is odd. For example, for $N = 4$ and $5$, $k^{\ast} = 2$ and for $N = 6$ and $7$, $k^{\ast} = 3$ and so on.} 
\label{f.1}
\end{figure}

\begin{figure}
%\centering
\includegraphics[width=0.9\columnwidth]{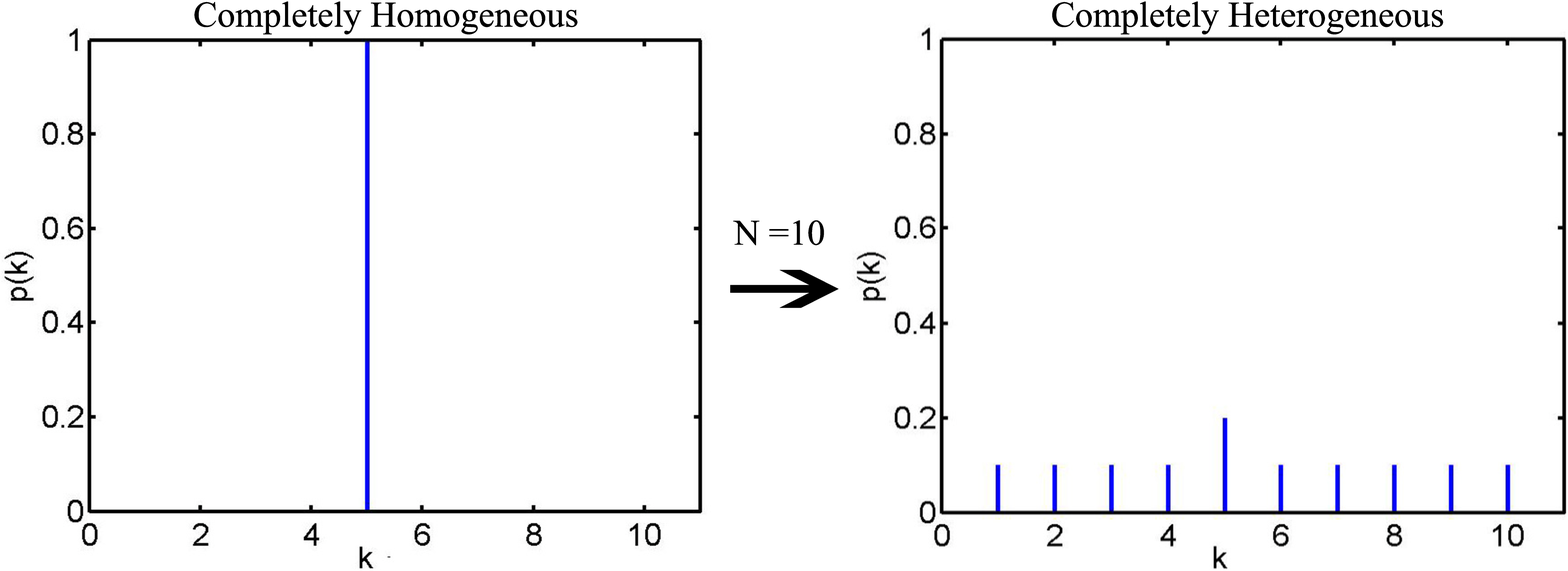}%
\caption{\label{f.2}Change in the degree distribution for a typical complex network as it is 
transformed from complete homogeneity to complete heterogeneity, for $N = 10$.} 
\label{f.2}
\end{figure}

If we analyse this measure closely, we find that it is basically different with respect to the 
earlier measures. The reason is that the measure proposed by Estrada is based on the Randic index 
\cite {randic} given by
 
\begin{equation}
R_{-1/2} = \sum_{i,j}(k_ik_j)^{-1/2}
  \label{eq:5}
\end{equation} 

Now, the Randic index was originally proposed \cite {xli} as a topological index under the name 
\emph{branching index} to measure the branching of Carbon atom skeletons of saturated Hydrocarbons. 
This index is so designed to get extremum value for the ``star'' structure which is the most heterogeneous 
branching structure and is bounded by values given by 

\begin{equation}
{\sqrt{N-1}} \leq R_{-1/2} \leq N/2
  \label{eq:6}
\end{equation}  
with the limiting values for the star structure and a regular lattice. To be specific, Estrada defines 
heterogeneity through an irregularity index for each pair of nodes $I_{ij}$, where $I_{ij} = 0$ if 
$k_i = k_j$ and $I_{ij} \rightarrow 1$ for $k_i = 1$ and $k_j \rightarrow \infty$. It is obvious that a 
measure based on this definition will be maximum for a ``star network'' of $N$ nodes compared to all other 
networks since there are $(N-1)$ connections with $I_{ij}$ having maximum value.

The above discussion makes it clear that Estrada's measure elegantly captures the structural aspect of 
heterogeneity associated with a complex network. This is also evident in the results given by the author. 
Out of all possible branching structures, the heterogeneity is maximum for the star structure. While 
the star network has $\rho_n = 1$, the values for networks with other topologies are much less with a 
typical SF network having $\rho_n \sim 0.1$. This measure is important in the context of real world 
networks with different topology and structure and can be used to classify such networks as 
shown by Estrada. 

Our focus here is the heterogeneity associated with the diversity in node degrees (analogous to the earlier  
attempts of heterogeneity) to propose a measure applicable to networks of all topologies. An important difference  
is that we use the frequencies of the node degrees, rather than $k_i$ directly, to define this measure. 
We show that, as the spectrum of $k$ values in the network increases, the 
value of the measure also increases correspondingly.  
We call the measure proposed here \emph {degree heterogeneity} in order to distinguish 
it from the measure in \cite {est2}.  Also, the two measures capture complimentary features 
of heterogeneity in a complex network. A network having high heterogeneity in one measure may not be so in 
the other measure and vice versa. For example, the star network is nearly homogeneous in our definition of 
heterogeneity, as shown below. It is also possible to correlate the robustness or stability of a network 
with the measure proposed here, with the SF networks having comparatively high value of heterogeneity. 
On the other hand, the star network is most vulnerable since disruption of just one node can destroy 
the entire network.   
To define the new measure, we require a network with a limiting value of heterogeneity to play a role similar to that 
of star network in the earlier measure. This network is presented in the next section. 

Finally, the heterogeneity measure that we define below can be shown to have direct correspondence with the 
entropy measure of a complex network \cite {bian}, characterized by the standard Shannon's measure of 
information $S$. In particular, this measure can be so adjusted to get the value zero for completely 
homogeneous networks and the value $S \rightarrow 1$ for the completely heterogeneous case as defined by 
us in this work. Though there are attempts to represent heterogeneity through entropy \cite {wu}, we prefer 
to view the two measures as two sperate aspects of a complex network. Entropy is usually associated with 
the rate at which a process or system (evolving) generates information. In the case of a network, each node 
can be considered as an information hub and links as channels for dissipating information. For the completely 
homogeneous network, no new information is generated while it tends to be maximum when the diversity in the 
node degrees is maximum. In this sense, entropy and heterogeneity are closely related and both have values 
normalized in the interval $[0,1]$. However, while entropy is a dynamic measure, heterogeneity is basically 
a static measure characterizing the structure and diversity of connections between the nodes and not 
directly concerned with the information transfer. That is why traditionally the two measures have been 
treated seperately, though the values of both for the extreme cases can be made identical.

Moreover, the measure that we propose below has the following advantages:

i) Only the degree distribution of the network is required to compute the heterogeneity in contrast to 
all the previous measures proposed so far.

ii) The specific condition that we apply for the completely heterogeneous network provides analytical values 
for heterogeneity in terms of network size.

iii) Based on the proposed measure, we are able to give a structural characterization index for a chaotic 
attractor through the construction of a complex network called recurrence network.  

\section{\label{sec:level1}COMPLETELY HETEROGENEOUS COMPLEX NETWORK}
Here we present what we consider as the logical limit of a completely heterogeneous network of $N$ nodes. 
The reader may find that this is an ideal case. Nevertheless, it helps to put the concept of heterogeneity 
of a complex network in a proper perspective.  
Consider an unweighted and undirected complex network of $N$ nodes, with all the nodes 
connected to the network having a degree of at least one. If all the nodes have the same degree $k$, the 
network is completely homogeneous with the degree distribution $P(k)$ being a $\delta function$ peaked at $k$. 

Let us now consider the other extreme where no two nodes have the same degree. The maximum possible degree for a 
node is $(N-1)$. Let the nodes be arranged in the ascending order of their degree. It is obvious that the 
$N^{th}$ node will have to take a degree equal to that of any one of the other nodes having degree from 1 to 
$(N-1)$. To find out what degree is possible for the $N^{th}$ node under the given condition,  
we start with taking small number of nodes as shown in  Fig.~\ref{f.1}, where we show $4$ 
different cases of $N$ ranging from $4$ to $7$. In each case, the $N^{th}$ node is represented as a 
\emph {pentagon shape} with its degree denoted as $k^{\ast}$. It is clear that if all the node degrees are to be  
different, there is only one possible value of $k^{\ast}$ for the $N^{th}$ node, which is ${{N} \over {2}}$ 
if $N$ is even and ${{(N-1)} \over {2}}$ if $N$ is odd.

\begin{figure}
%\centering
\includegraphics[width=0.9\columnwidth]{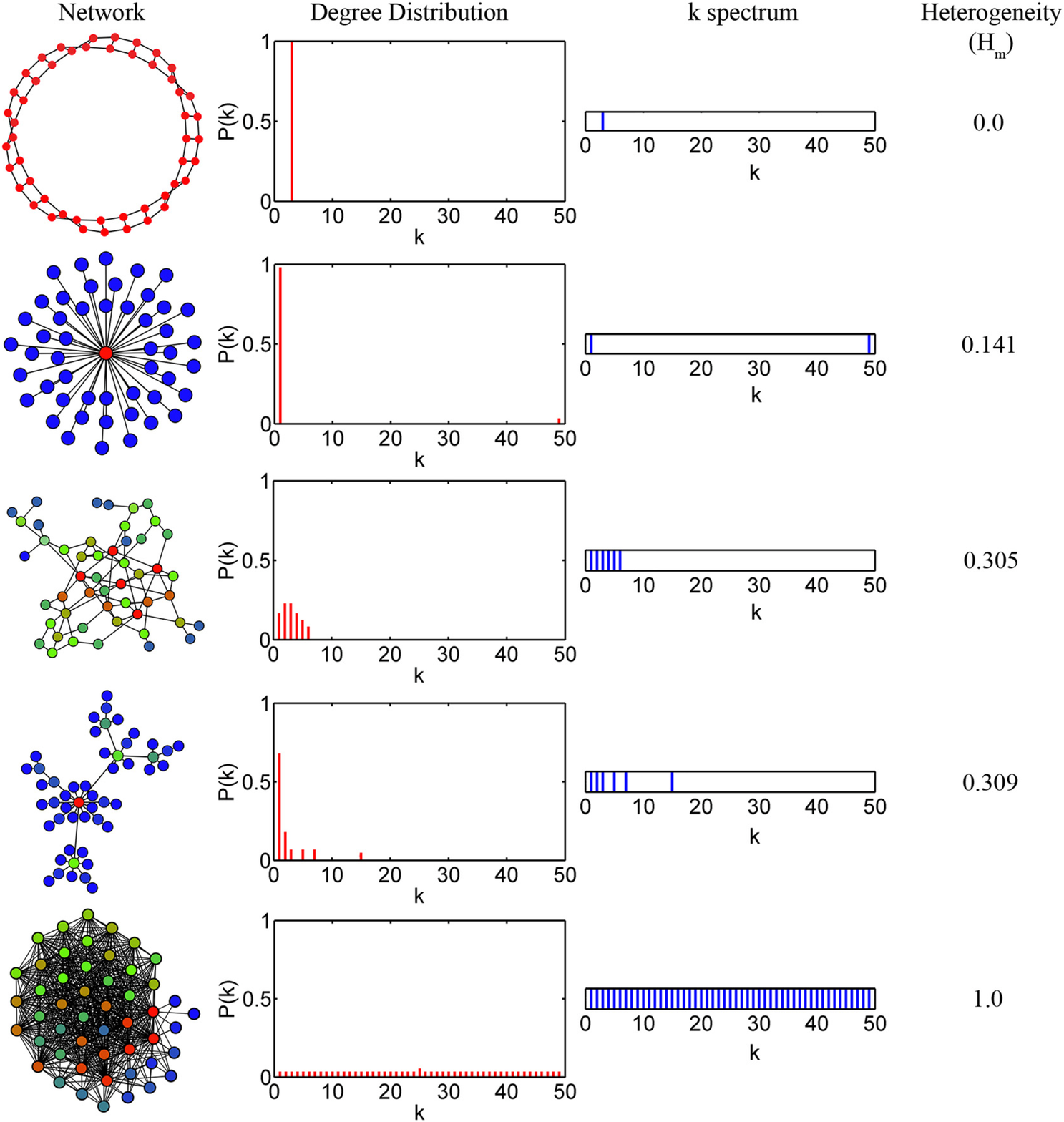}%
\caption{\label{f.3}A snapshot of different types of complex networks in the increasing order of their 
heterogeneity($H_m$), taking $N = 50$ in all cases. From top to bottom, the nature of the network varies 
from completely homogeneous, star, RG, SF and finally to completely heterogeneous network. 
The degree distribution and $k$ spectrum are also 
shown for each case to indicate that $H_m$ is a measure of the diversity in the node degrees.} 
\label{f.3}
\end{figure}

We now give a simple argument that this result is true in general for any $N$. The degree of node 1 is $1$ 
which means that it is connected only to the node with degree $(N-1)$. That is, node 1 is not connected to 
$N^{th}$ node. Node 2 is connected only to two nodes with degree $(N-1)$ and $(N-2)$ and hence it is also 
not connected to node $N$. By induction, one can easily show that the $r^{th}$ node is connected only to 
nodes with degree $(N-1)$, $(N-2)$,.......$(N-r)$. Suppose $N$ is even. When $r = {{N} \over {2}}$, this 
node is connected to nodes with degree from $(N-1)$ to ${{N} \over {2}}$. To avoid self loop, this node should be 
connected to node $N$. Thus all nodes with higher degree from ${{N} \over {2}}$ to $(N-1)$ are connected to 
node $N$ whose degree becomes ${{N} \over {2}}$. By a similar argument, one can show that the degree of 
$N^{th}$ node is ${{(N-1)} \over {2}}$ if $N$ is odd.

Let us now consider the degree distribution $P(k)$ of this completely heterogeneous network. All the nodes have 
different $k$ values and only two nodes share the same $k$ value, $k^{\ast}$. One can easily show that:
\begin{eqnarray*}
P(k) = P_0 = {{1} \over {N}},   (k \neq k^{\ast})   \nonumber \\
P(k) = {{2} \over {N}},  (k = k^{\ast})      \nonumber  \\
\end{eqnarray*}    
Our definition of heterogeneity  is derived in such a way that this network has maximum heterogeneity, 
which is done in the next section.

\section{\label{sec:level1}A NEW MEASURE OF DEGREE HETEROGENEITY}
It is very well accepted that a network of $N$ nodes with all nodes having equal degree $k$ is a completely 
homogeneous network with $P(k)$ being a $\delta function$ centered at $k$. The value of $k$ can be anything in 
the range $2 \leq k \leq (N-1)$ and all these networks have heterogeneity measure zero, for any $N$. In 
principle, the heterogeneity of a network should measure the diversity in the node degrees with respect to a 
completely homogeneous network of same number of nodes. All the measures defined so far in the literature 
directly use the $k$ values present in the network for computing the heterogeneity measure. Here we argue 
that a much better candidate to define such a measure is $P(k)$ rather than $k$. Since $P(k)$ is a 
probability distribution, as the spectrum of $k$ values increase, the value of $P(k)$ gets shared between 
more and more nodes with the condition $\sum_k P(k) = 1$. In other words, this variation in $P(k)$ 
reflects the diversity of node degrees and hence the heterogeneity of the network. A typical variation of 
$P(k)$ as the network changes from complete homogeneity to complete heterogeneity is shown in 
Fig.~\ref{f.2}. Note that for RGs, this variation in $P(k)$ is with respect to $P(<k>)$, with $<k>$ 
being the average degree, while for SF networks, it is with respect to $P(k_{min})$. 

To get the heterogeneity measure, we first define a heterogeneity index $h$ for a network of $N$ nodes 
as the variance of $P(k)$ with respect to the peak value corresponding to the completely homogeneous 
case:
\begin{equation}
h^2 = {{{1} \over {N}} \sum_{k_{min}}^{k_{max}} (1 - P(k))^2}, P(k) \neq 0
  \label{eq:7}
\end{equation} 
The condition implies that the summation is only over $k$ values for which $P(k) \neq 0$. For a 
completely homogeneous network, $P(k)$ is non zero only for one value of $k$, say $k^{\ast}$, and 
$P(k^{\ast}) = 1$, making $h = 0$, for all $N$. 

We now consider the other extreme of completely heterogeneous case. From the results in the previous 
section for the completely heterogeneous case, we have  
\begin{equation}
h_{het}^2 = {{{1} \over {N}} \sum_{k=1}^{(N-1)} (1 - P(k))^2}
  \label{eq:8}
\end{equation} 
Putting the values of $P(k)$ and simplifying, we get 
\begin{equation}
h_{het}^2 = {1 - {{3} \over {N}} + {{N+2} \over {N^3}}}
  \label{eq:9}
\end{equation}
This is the maximum possible heterogeneity measure for a network of $N$ nodes. For large $N$,  
as a first approximation, we have 
\begin{equation}
h_{het} \approx {\sqrt{{1 - {{3} \over {N}}}}} 
  \label{eq:10}
\end{equation}
For finite $N$, its value is $< 1$ and as $N \rightarrow \infty$, $h_{het} \rightarrow 1$. 
To define the heterogeneity measure $(H_m)$ for 
a network, we normalize the heterogeneity index of the network with respect to the completely 
heterogeneous network of same number of nodes to get the value in the unit interval $[0,1]$:
\begin{equation}
H_m = {{h} \over {h_{het}}}
  \label{eq:11}
\end{equation}
If $N$ is sufficiently large, say $N > 1000$ as is the case for most practical networks,  $h_{het} \sim 1$ 
and $H_m \approx h$. 

\begin{figure}
%\centering
\includegraphics[width=0.9\columnwidth]{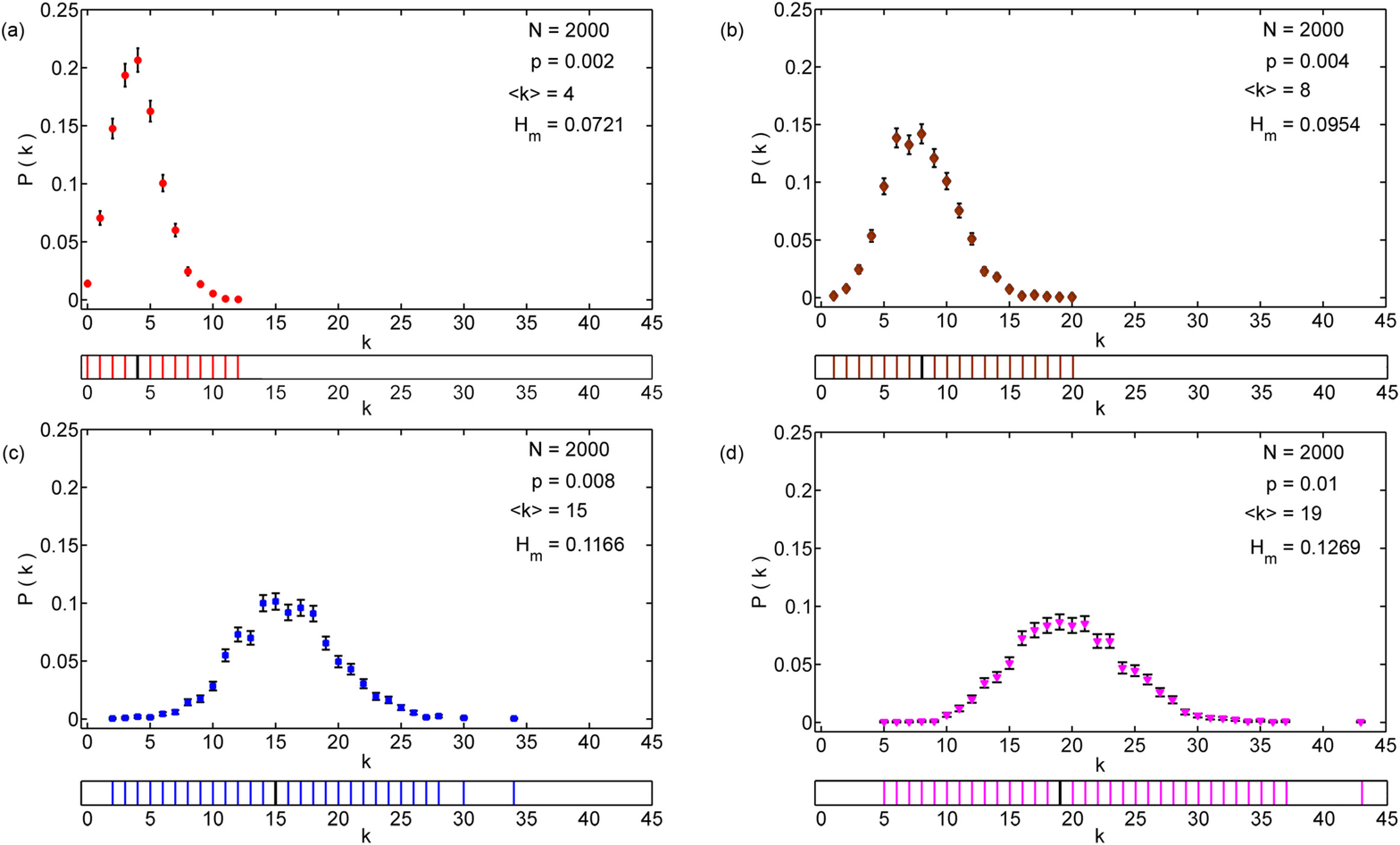}%
\caption{\label{f.4}Degree distribution of E-R networks (RGs) for four different $p$ values with $N$ 
fixed at $2000$. The $k$ spectrum is shown below the degree distribution. The value of $H_m$ and $<k>$ are 
also indicated in each case. Note that $H_m$ varies directly with the degree diversity or the spectrum 
of $k$ values in the network.} 
\label{f.4}
\end{figure}

\begin{figure}
%\centering
\includegraphics[width=0.8\columnwidth]{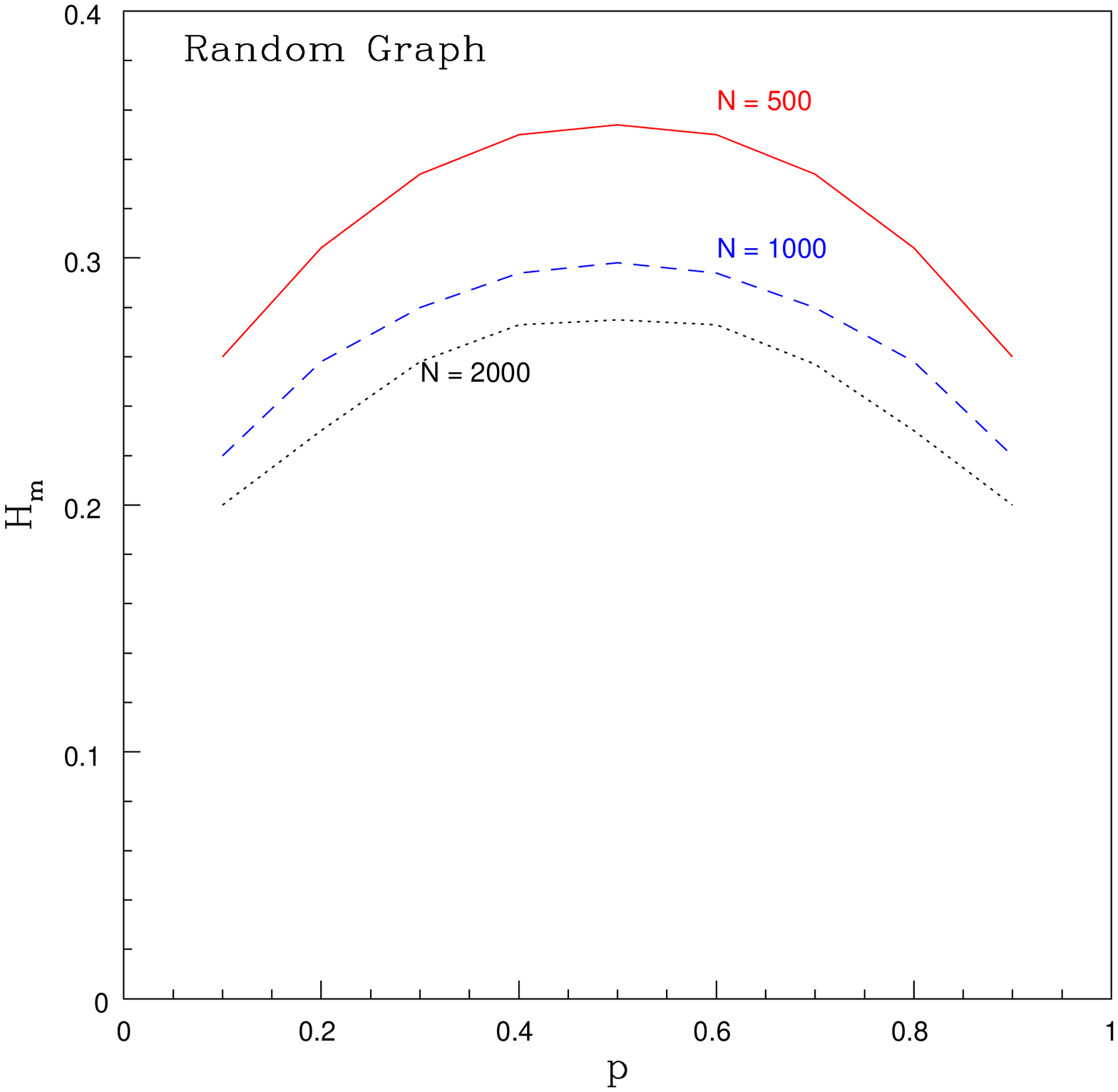}%
\caption{\label{f.5}Variation of $H_m$ with $p$ for RGs for a fixed value of $N$, 
as shown. We expect the profile for a higher $N$ value to be within that of a lower $N$ as $H_m$ 
decreases with $N$ for any fixed $p$.} 
\label{f.5}
\end{figure}

We note the following features regarding $H_m$:

i) It is defined here for unweighted and undirected complex networks and represents a unique measure 
applicable to any network independent of the topology or degree distribution and increases with the 
diversity in the node degrees. 

ii) However, certain topologies have inherent limitations in diversity. For example, $H_m$ for a star 
network is very close to zero and hence the star network is nearly homogeneous in our definition. 
This is because, in the star topology, the degree of only one node is different from the rest of the nodes. 

iii) Since we use the counts of the node degrees rather than directly $k_i$ to find $H_m$, we cannot express 
the measure in terms of the elements of the Laplacian matrix, as some authors have done.

iv) For two networks of the same size $N$ independent of the topology, the measure we propose has a 
direct correspondence with the degree diversity in the network. To show this 
explicitly, we present the \emph{k spectrum}, the spectrum of $k$ values in the network in the form of a 
discrete line spectrum. In Fig.~\ref{f.3}, we compare some standard networks in the increasing order of 
their $H_m$, taking $N = 50$. For each network, we show the degree distribution (as histogram), the 
$k$ spectrum and the value of $H_m$. Note that, of different topologies, the SF 
network is the most heterogeneous. Here the star network has a reasonably high value of $H_m$ since 
$N$ is only $50$. We also show the completely heterogeneous network with $H_m = 1$, for comparison. 

v) The heterogeneity index $h$ is defined as a measure normalized with respect to the size of the network 
$N$. For large $N$, since $h \sim H_m$, the measure $H_m$ can also 
be used to compare the heterogeneities of two networks even if $N$ is different. This is especially important 
for real world networks where $N$ varies from one network to another, as discussed in \S VI. However, a 
network with larger $N$ generally tends to have lower $H_m$ since, to keep the same heterogeneity, the 
range of non zero $k$ values should also increase correspondingly. In other words, a network with $100$ 
nodes attains complete heterogeneity if the $k$ values range from $1$ to $99$ whereas, to attain complete 
heterogeneity for a network of $1000$ nodes, the $k$ values should range from $1$ to $999$. 

The above result also implies that for any network that is evolving or growing, for example the SF network where the 
nodes are added with preferential attachment \cite {alb3}, the value of $H_m$ generally keeps on decreasing 
with increasing $N$. In the next section, we numerically study the variation of $H_m$ with different 
network parameters for various synthetic networks.

\section{\label{sec:level1}DEGREE HETEROGENEITY OF SYNTHETIC NETWORKS}
In this section, we analyze $3$ different classes of complex networks, namely, the RGs of Erdos-Renyi, 
the SF networks and the networks derived from the time series of chaotic dynamical systems, called 
recurrence networks (RNs) whose details are discussed in \S V.C.

\subsection{Classical random graphs}
For RGs, the degree distribution is Poissonian centered around an average degree $<k> \equiv pN$ where 
$p$ is the probability that two nodes in the network is connected. In Fig.~\ref{f.4}, we show the degree 
distribution and the $k$ spectrum for RGs of $4$  different $p$ values with $N$ fixed at $2000$. The 
values of $H_m$ for all these networks are also shown. The main result here is that the value of 
$H_m$ increases correspondingly with the range of $k$ values for a fixed $N$. 

We next consider how $H_m$ depends on $p$ and $N$, the two basic parameters of the RG. The effect of 
changing $p$ for a fixed $N$ as well as changing $N$ for a fixed $p$ are to shift the average $k$ 
value of the nodes in the RG. Since the degree distribution is approximately Gaussian for large $N$, 
the spectrum of 
$k$ values depends directly on the variance of the Gaussian profile. As $p$ increases from zero for a 
fixed $N$, the spectrum of $k$ values and hence $H_m$ increase correspondingly. Due to the obvious 
symmetry of the network with respect to the transformation $p \rightarrow (1-p)$,  
as $p$ increases beyond $0.5$, $H_m$ starts decreasing. Thus the maximum value of $H_m$ is 
obtained for $p = 0.5$ for any fixed $N$.  
On the other hand, by increasing $N$ for any fixed $p$, one expects the 
Gaussian profile of the degree distribution to become sharper, thus decreasing $H_m$.  These results are compiled in 
Fig.~\ref{f.5} for three values of $N$. Note that the minimum $p$ value that can be used for $N = 500$ 
is $0.004$ and this decreases as $N$ increases. In the figure, we show the results starting from $p = 0.1$. 
Higher values of $N$ would involve very large computer 
memory requirements for large $p$. However, we 
have checked the variation of $H_m$ with $N$ for smaller $p$ values, say $0.005$ and $0.01$, for 
$N$ up to $5000$ and have found that the decrease is approximately exponential.  

\begin{figure}
%\centering
\includegraphics[width=0.9\columnwidth]{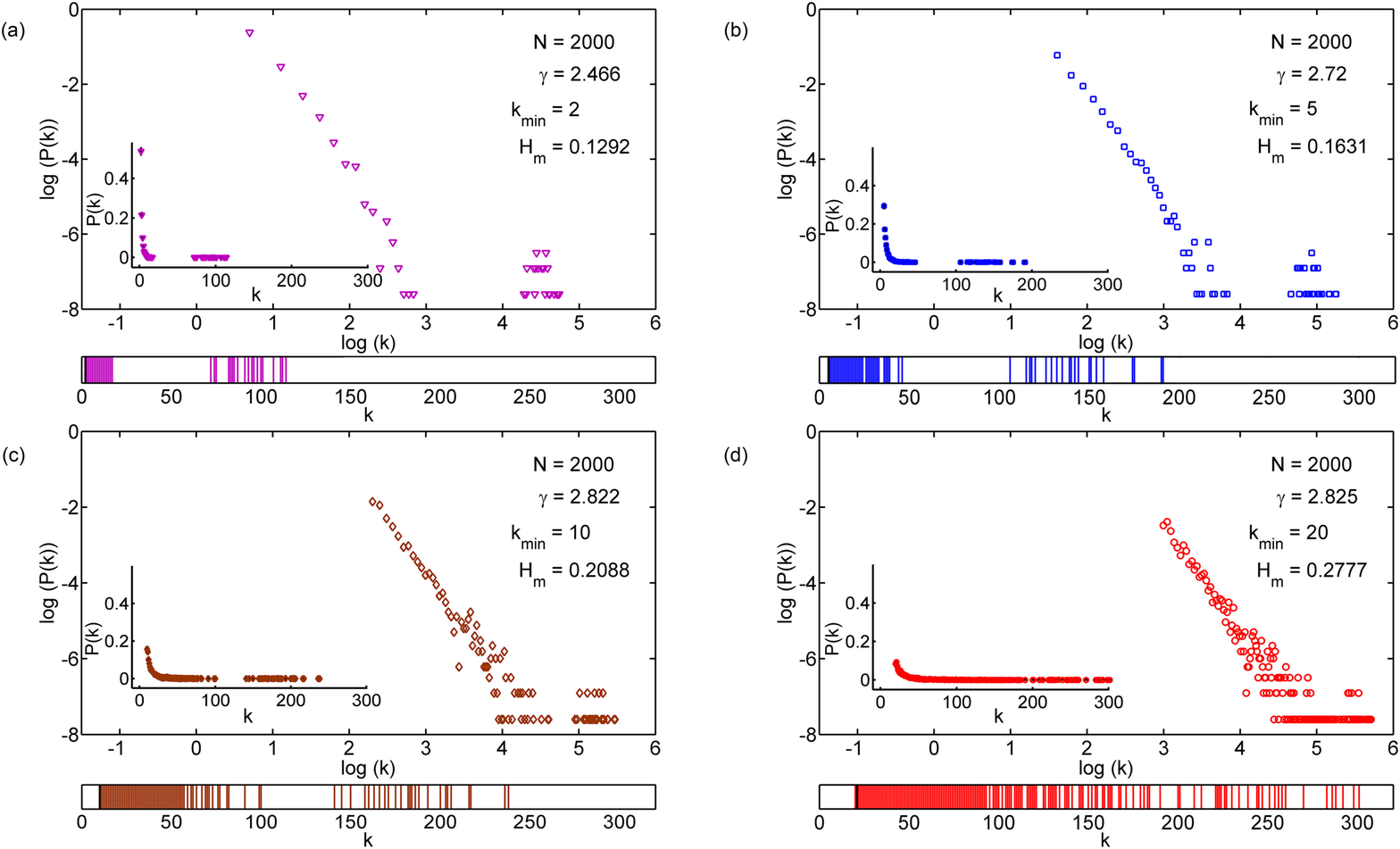}%
\caption{\label{f.6}Degree distributions (inset) and the distributions in log scale along with the 
$k$ spectrum for synthetic SF networks with four different values of $\gamma$ and $N$ fixed at $2000$. 
In all cases, the values of 
$H_m$ and the minimum degree $k_{min}$ of the network are also shown. As the spectrum of $k$ values increases, 
$H_m$ increases correspondingly for a fixed $N$. Note that there appears to be a second peak with a gap in 
the distribution for $log k > 4$. This is size dependent effect due to the presence of many $k$ values 
having $P(k)$ very close to zero. It is also evident from the $k$ spectrum shown below each distribution. 
For example, in the case $(d)$ where the $k$ spectrum is almost continuous without a visible gap, the scaling 
becomes more evident. } 
\label{f.6}
\end{figure}

\begin{figure}
%\centering
\includegraphics[width=0.9\columnwidth]{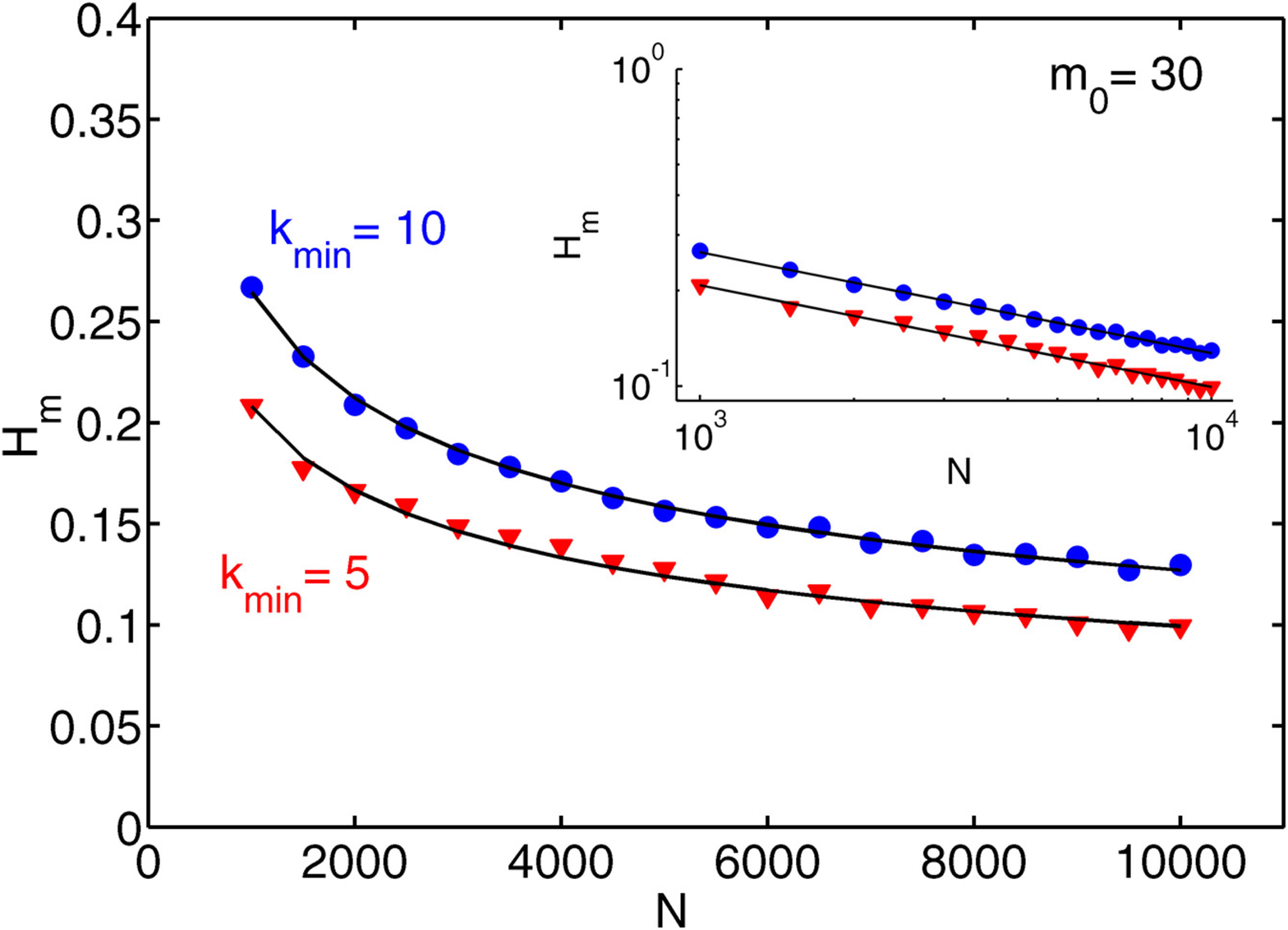}%
\caption{\label{f.7}Variation of $H_m$ with $N$ for synthetic SF networks with two different values of $k_{min}$. 
The same variation is shown in the inset in log scale indicating a clear power law in both cases.} 
\label{f.7}
\end{figure}

\subsection{Scale free networks}
For SF networks, the degree distribution obeys a power law $P(k) \propto k^{-\gamma}$. To construct the SF 
network synthetically, we use the basic scheme proposed by Barabasi et al. \cite {barab}.  
In this scheme, we start with a small number of initial nodes denoted as $m_0$. As a new node is 
connected, a fixed number of edges, say $m$, is added to the network. This number represents the 
minimum number of node degree, $k_{min}$, in the network. The new edges emerging at node creation are 
distributed according to the preferential attachment mechanism. 
The two parameters, $m_0$ and $k_{min}$, 
determines the value of $\gamma$ as the network evolves. We have constructed SF networks of different 
$\gamma$ by changing both $m_0$ and $k_{min}$. 

In Fig.~\ref{f.6}, we show the degree distribution and the corresponding $k$ spectrum for SF networks 
with four different $\gamma$ and $k_{min}$, with $N$ fixed at $2000$. We find that the $k$ spectrum and 
hence the value of $H_m$ depend directly on $k_{min}$ as can be seen from the figure. In other words, 
for a SF network of fixed $N$, $H_m$ increases as the value of $k_{min}$ increases. The variation is 
approximately linear for $k_{min}$ in the range $1$ to $10$. More interesting is the variation of 
$H_m$ with $N$ for a fixed $k_{min}$. In Fig.~\ref{f.7}, we show the variation of $H_m$ as $N$ 
increases from $1000$ to $10000$ for two different SF networks with $k_{min} = 5$ and $10$. This 
variation is also shown in the inset in a log scale in the same figure indicating that $H_m$ varies as  
$H_m \propto N^{-\rho}$, where the value of $\rho$ is found to be $0.3144$ for $k_{min} = 5$ and 
$0.3290$ for $k_{min} = 10$ for the given range of $N$ values.  However, we do not claim that this 
variation is, in general, a power law since we have only tested a limited range of $N$ values. This 
needs to be explicitly tested with other alternatives with a wider range of $N$ values.

\subsection{Recurrence networks}
Recently, a new class of complex networks has been proposed for the characterization of the structural 
properties of chaotic attractors, called the recurrence networks (RNs)\cite {don1,don2}. They are constructed 
from the time series of any one variable of a chaotic attractor. From the single scalar time series, 
the underlying attractor is first constructed in an embedding space of dimension $M$ using the time 
delay embedding \cite {spr} method. Any value of $M$ equal to or greater than the dimension of the 
attractor can be used for the construction of the attractor. The topological and the structural 
properties of this attractor can be studied by mapping the information inherent in the attractor to a 
complex network and analyzing the network using various network measures.

To construct the network, an important property of the trajectory of any dynamical system is made use of, 
namely, the recurrence \cite {eck}. By this property, the trajectory tends to revisit any infinitesimal 
region of the state space of a dynamical system covered by the attractor over a certain interval of time. 
To convert the attractor to a complex network, 
one considers all the points on the embedded attractor as nodes and two nodes $\imath$ and $\jmath$ are 
considered to be connected if the distance $d_{ij}$ between the corresponding points on the attractor in 
the embedded space is less than or equal to a recurrence threshold $\epsilon$. Selection of this 
parameter is crucial in getting the optimum network that represents the characteristic properties of the 
attractor. The resulting complex network 
is the RN which, by construction, is an unweighted and undirected network. The adjacency matrix 
$\mathcal A$ of the RN is a binary symmetric matrix with elements $A_{ij} = 1$ (if nodes $\imath$ 
and $\jmath$ are connected) and $0$ (otherwise). More details regarding the construction of the RN can be 
found elsewhere \cite {dong,rj}. Here we follow the general framework recently proposed by us 
\cite {rj} to construct the RN from time series. 

\begin{figure}
%\centering
\includegraphics[width=0.9\columnwidth]{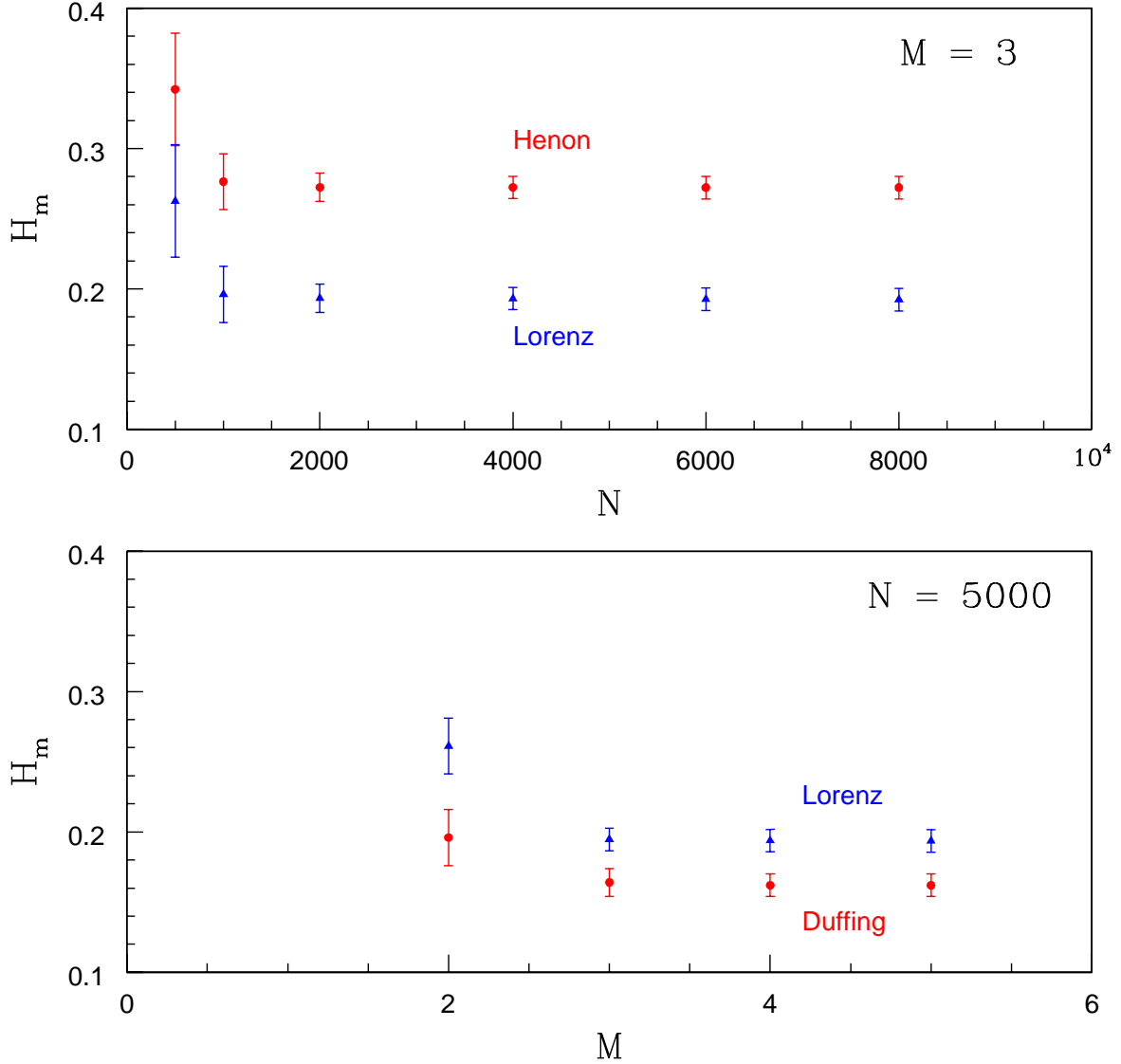}%
\caption{\label{f.8}Top panel shows the variation of $H_m$ with $N$ for RNs constructed from Lorenz 
(filled triangle) and Henon (filled circle) attractor time series. Bottom panel 
shows the variation of $H_m$ with $M$ for fixed $N$ for Lorenz (filled triangle) and Duffing (filled 
circle) attractors. In both graphs, the error bar comes 
from the standard deviation of values for $H_m$ computed from time series with ten different initial conditions.} 
\label{f.8}
\end{figure}

For generating the time series, we use the equations and the parameter values given in \cite {spr} for all 
chaotic systems. For continuous systems, we have used the sampling rate $0.05$ for generating the time 
series. The time delay used for embedding is the first minimum of the autocorrelation. 
We first study how the value of $H_m$ varies with the number of nodes $N$ for RNs. 
In Fig.~\ref{f.8} (top panel), we show the results for 
the Lorenz attractor and the Henon attractor. It is evident that for large value of $N$, $H_m$ converges 
to a finite value. We have checked and verified that this is true for other low dimensional chaotic 
attractors as well. It is found that once the basic structure of the attractor is 
formed, the value of $H_m$ remains independent for further increase in $N$. In other words, the range of 
$k$ values  increases with $N$ to keep the value of $H_m$ approximately constant. This result also follows from the 
statistical invariance of the degree distribution of the RN as has already been shown \cite {rj}. 

Next, we consider the variation of $H_m$ with embedding dimension $M$. This is also shown in Fig.~\ref{f.8} 
(bottom panel) for two standard chaotic attractors. It is clear that the value of $H_m$ converges for 
$M \geq 3$ in both cases. This is because, the fractal dimension of both these attractors are $< 3$. 
We have already shown \cite {rj} that the degree distribution of the RN from any chaotic attractor 
converges beyond the actual dimension of the system. 
Thus, $H_m$ turns out to be a unique measure for any chaotic attractor independent of both $M$ and $N$.  

From the construction of RNs, the range of connection between two nodes is limited by the recurrence threshold 
$\epsilon$. Hence the degree of a node in the RN and the probability density around the corresponding 
point over the attractor are directly related. For example, for the RN from a random time series, 
every node has degree close to the average value $<k>$ since the probability density over the attractor 
is approximately the same. One can show that the degree distribution of the RN from a random time 
series is Gaussian for large $N$. Thus, the $k$ spectrum of the RN is indicative of the range in the 
probability density variations over the attractor, which in turn, is characteristic of the structural 
complexity of the attractor. 

We have already shown that the measure $H_m$ proposed here is indicative of the diversity in the 
$k$ spectrum. Moreover, it is found to have a specific value for a given attractor independent of $M$ and $N$. 
It is well known that the statistical measures derived from the RNs characterize the structural 
properties of the corresponding chaotic attractor. In particular, since every point on the attractor is 
converted to a node in the RN and the local variation in the node degree is a manifestation of the 
variation in the local probability density over the attractor, the measure 
$H_m$ can serve as a single index to quantify the structural complexity of a 
chaotic attractor through RN construction. In Table I, we compare the values of $H_m$ for RNs constructed 
from several standard chaotic attractors. In all cases, the saturated value of $H_m$ 
converged upto $M = 5$ is shown. In each case, ten different RNs are constructed changing the initial 
conditions and the average is shown with standard deviation as the error bar.  The results indicate that 
that among the continuous systems compared, the Lorenz attractor is structurally the most complex while in 
the case of 2D discrete 
systems, Lozi attractor is found to be the most diverse in terms of the probability density variations. 

Finally, it will also be interesting to see how the value of $H_m$ is affected by adding noise to the 
chaotic time series. To test this, we generate data adding different percentages of noise to Lorenz 
data. When the value of $H_m$ is computed, it is found that the value reduces systematically with the 
increase in the noise percentage and approaches the value of noise for a noise level $> 50 \%$. This 
result clearly indicates that the measure will be useful for analysing the real world data.
 
The value of $H_m$ for random time series with $N = 2000$ and  $M = 3$ is found to be $0.084 \pm 0.012$. 
To get a comparison with the values of conventional complex networks, we compute $H_m$ for a RG with 
$p = 0.0035$ that gives the same $<k>$ as that of the RN from random time series and a typical SF 
network with $\gamma = 2.124$ with $N = 2000$ in both cases.  The average of ten different simulations is 
taken. We find that $H_m = 0.087 \pm 0.012$ for the RG which is exactly same as the RN from random 
time series and $H_m = 0.114 \pm 0.06$ for the SF network.  

\begin{table}[h]
\centering
\begin{tabular}{|l||c|c|c|c|c|c|c|}
\hline
\emph{System} & Lorenz & R\"ossler & Duffing & Ueda & Henon & Lozi & Cat Map  \\
\hline

$H_m$ & $0.1942 \pm 0.056$ & $0.1874 \pm 0.042$ & $0.1686 \pm 0.058$ & $0.1662 \pm 0.038$ & $0.2582 \pm 0.044$ & $0.2744 \pm 0.072$ & $0.1280 \pm 0.048$   \\

\hline
\end{tabular}
\caption{Comparison of $H_m$ for several standard chaotic attractors.}
\label{tab:1}
\end{table}

\begin{figure}
%\centering
\includegraphics[width=0.8\columnwidth]{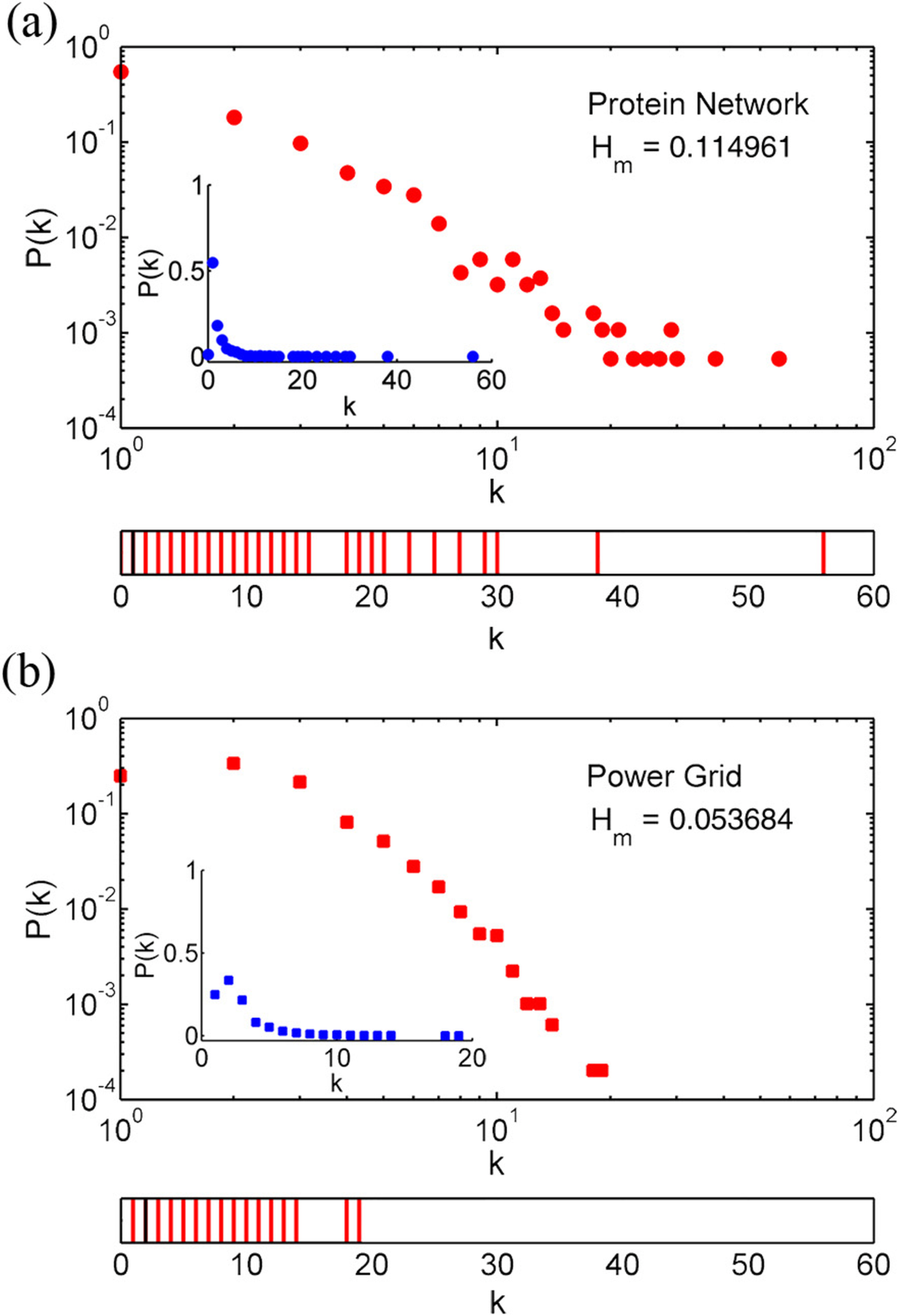}%
\caption{\label{f.9}Degree distribution and the $k$ spectrum of the protein interaction network are 
shown in the top panel. In the bottom panel, the same for the network of Western Power Grid.} 
\label{f.9}
\end{figure}

\section{\label{sec:level1}REAL WORLD NETWORKS AND POSSIBLE EXTENSION TO WEIGHTED NETWORKS}
So far, we have been discussing the degree heterogeneity measure of synthetic networks of different topologies. 
In this section, we consider some unweighted and undirected complex networks from the real world and see 
what information regarding the degree heterogeneity of such networks can be deduced using the proposed measure. 
We use data on networks from a cross section of fields, such as, biological, technological and social 
networks.  In Fig.~\ref{f.9}, we show the degree distribution and $k$ spectrum of two such networks. 
In Table II, we compile the details of these networks and the values of $H_m$ computed by us for each. 

\begin{table}[h]
\centering
\begin{tabular}{||l|c|c|c||}
\hline
\emph{System} & \emph{Reference} & $N$ & $H_m$ \\
\hline
\hline

US Power Grid & http://cdg.columbia.edu/cdg/datasets & 4941 & 0.056 \\
              & \cite{wat}                           &      &        \\
 
& & & \\

Protein Interaction & www3.nd.edu/~networks/resources.html & 1846 & 0.1182 \\
                    & \cite{jeo}                           &      &        \\

& & & \\

Budding Yeast & math.nist.gov/ & 2353 & 0.1542 \\
              & \cite{sun}     &      &        \\

& & & \\

US Patent Citation & https://snap.stanford.edu/data & 7253 & 0.027 \\

& & & \\

Dolphin Interaction & https://snap.stanford.edu/data & 62 & 0.386  \\
                    & \cite{luss}                    &    &         \\
\hline
\hline
\end{tabular}
\caption{Comparison of the degree heterogeneity measure of five real world networks}
\label{tab:2}
\end{table}

Since we have to restrict to the case of unweighted and undirected networks, we could use only a small 
subset from the very large variety of real world networks that are mostly weighted or directed.  
To extend the measure to directed networks, one has to consider the in-degree and out-degree 
distributions and find the heterogeneity separately. In order to generalise the measure to weighted 
networks, the distribution of the weight or strength of the nodes in the network \cite {yoo,barr}, 
rather than the simple degree distribution is to be considered and define the measure accordingly. For example, 
for unweighted and undirected networks, all the links are equivalent and hence the degree of $\imath^{th}$ 
node $k_i$ is just the sum of the links connected to node $\imath$.  
On the other hand, for weighted networks, each link is associated 
with a weight factor $w_{ij}$ and hence the degree $k_i$ should be generalised to the sum of the weights 
of all the links attached to node $\imath$:
\begin{equation}
s_i = \sum_j w_{ij}
  \label{eq:12}
\end{equation}
Thus the degree distribution needs to be generalised to the \emph{strength distribution} $P(s)$, which is 
the probability that a given node has a strength equal to $s$ \cite {ba,lat}. The equation for heterogeneity 
for weighted networks can be modified accordingly. However, it should be noted that the weight factors are 
assigned based on different criteria depending on the specific system or interaction the network tries to 
model. Modifying the measure by incorporating the specific aspects of interaction, the measure itself 
becomes network specific. The measure that we propose here is independent of such details and is 
representative of only the diversity of node degrees in a network, determined completely by the simple 
degree distribution. 
  
\section{\label{sec:level1}CONCLUSION}
Complex networks and the network based quantifiers have become useful tools for the analysis of many real world 
phenomena. Physical, biological and social interactions are increasingly being modeled and characterized through 
the language of complex network. 
An important measure for the characterization of any complex network is its heterogeneity measured in terms of 
the diversity of connection reflected through its node degrees. Here we introduce a measure to quantify this 
diversity which is applicable to networks of different topologies. This measure is minimum (equal to 
zero) for a completely homogeneous network with all $k_i \equiv k$. To get the upper bound for the measure, 
we consider the logical limit of heterogeneity possible in a network of $N$ nodes where nodes of degree 
varying from 1 to $(N-1)$ are present, whose heterogeneity is normalised as 1. While considering this network 
of limiting heterogeneity, we also prove  that the degree that is repeated (or shared by two nodes) 
is $N/2$ if $N$ is even and ${(N-1)} \over {2}$ if $N$ is odd.  

Also, the measure that we propose here uniquely quantifies the diversity in the node degrees in the network 
which is characteristic of the type and range of interactions the network represents. The diversity also depends on the 
topology of the resulting network. For example, for RGs, the diversity is limited since most degrees are 
centered around the average value $<k>$ while the SF networks are comparatively more diverse due to the 
presence of hubs. The proposed measure can quantify this diversity in the node degrees   
irrespective of the topology of the network. 

By applying the proposed measure, we compute the heterogeneity of various unweighted and undirected networks, 
synthetic as well as real world. We study numerically how the heterogeneity varies for RG with respect to the two 
parameters $p$ and $N$, while for SF networks the variation of heterogeneity with respect to $\gamma$ 
as well as $N$ are analysed. To illustrate the practical relevance of the measure, we 
analyse the RNs constructed from the time series of chaotic dynamical systems and  highlight its utility as a 
quantifier to compare the structural complexities of different chaotic attractors. As has already shown by us 
\cite {rj}, the nonlinear character and chaotic dynamics underlying the time series can be distinguished from the 
RNs through the usual characteristic measures of complex networks like CC or CPL. 
However the subtle differences between the degree distributions of RNs from different chaotic systems are not clearly 
evident from their CC or CPL. We find these can be quantified uniquely using the proposed measure of degree heterogeneity.

{\bf Data accessibility}: We use the open source software GEPHI \emph{www.gephi.org/} for the construction of all 
networks. All the codes used for computing the heterogeneity and other network measures are available at 
\emph {https://sites.google.com/site/kphk11/home.} 

{\bf Author contributions}: The initial idea for the paper started from the discussions of RJ and KPH. 
The codes for computations of network measures were developed in association with RM. The interpretation 
of the results and the expert guidance for the manuscript were finalised after a series of discussions 
with GA. All the authors gave the final approval of the manuscript.  

{\bf Competing interests}: The authors have no competing interests. 

{\bf Funding}: RJ, KPH and RM acknowledge the financial support from the Science and Engineering Research  
Board (SERB), Govt. of India in the form of a Research Project No. SR/S2/HEP-27/2012.

{\bf Acknowledgements}: The authors thank one of the anonymous Referees for a thorough review of the 
manuscript and especially for pointing out the correspondence of our heterogeneity measure with the 
entropy measure of complex networks. 
RJ and KPH acknowledge the computing facilities in IUCAA, Pune.

\end{document}